# Improving GGH Public Key Scheme Using Low Density Lattice Codes

Reza Hooshmand
Department of Electrical Engineering,
Science and Research Branch, Islamic Azad University, Tehran, Iran
r.hooshmand@srbiau.ac.ir

*Abstract*—**Goldreich-Goldwasser-Halevi (GGH) public key cryptosystem is an instance of lattice-based cryptosystems whose security is based on the hardness of lattice problems. In fact, GGH cryptosystem is the lattice version of the first code-based cryptosystem, proposed by McEliece. However, it has a number of drawbacks such as; large public key length and low security level. On the other hand, Low Density Lattice Codes (LDLCs) are the practical classes of lattice codes which can achieve capacity on the additive white Gaussian noise (AWGN) channel with low complexity decoding algorithm. This paper introduces a public key cryptosystem based on LDLCs to withdraw the drawbacks of GGH cryptosystem. To reduce the key length, we employ the generator matrix of the used LDLC in Hermite normal form (HNF) as the public key. Also, by exploiting the linear decoding complexity of the used LDLC, the decryption complexity is decreased compared with GGH cryptosystem. These increased efficiencies allow us to use the bigger values of security parameters. Moreover, we exploit the special Gaussian vector whose variance is upper bounded by the Poltyrev limit as the perturbation vector. These techniques can resist the proposed scheme against the most efficient attacks to the GGH-like cryptosystems.**

## I. INTRODUCTION

Code and lattice-based cryptography are two of the most promising candidates to post quantum cryptography. It is believed that code and lattice-based cryptosystems are able to resist the attacks performed by quantum computers. In fact, there is no known quantum algorithm allows attacking them significantly faster than classical algorithms. The security provided by such cryptosystems is based on the difficulty of some classical problems related to coding theory and lattices, respectively [1]. Up to now, various public key code and lattice-based cryptosystems have been introduced and developed. One of the instances of lattice-based public key cryptosystems is proposed by Goldreich, Goldwasser and Halevi, known as GGH cryptosystem [2]. In fact, GGH public key scheme is the lattice analog of the McEliece cryptosystem [3] in which Goppa codes are replaced by lattices. The GGH scheme has some advantages, such as [4]: (1) it seems to be more secure than RSA and ElGamal encryption schemes against quantum computers; (2) it has a natural signature scheme; (3) unlike NTRU cryptosystem [5], the secret key of the GGH scheme cannot be obtained by solving the shortest vector problem (SVP). However, it has a number of weaknesses as follows [6]: (1) because of special form of the perturbation vector employed in the GGH scheme, the problem of decrypting ciphertext can be reduced to easier problem and thus partial information on plaintext can be recovered; (2) the GGH cryptosystem is insecure against adaptive chosen ciphertext attacks (CCA2); (3) its key length is very large.

On the other hand, Low Density Lattice Codes (LDLCs) are an efficient class of lattice codes which can achieve capacity on the additive white Gaussian noise (AWGN) channel [7]. LDLCs have the nonsingular generator matrix $G$, i.e., $\det(G) = 1$, and their parity check matrix $H = G^{-1}$ is restricted to be sparse. The sparsity of $H$ in the LDLCs is utilized to develop a low complexity iterative decoding algorithm. The main goal of this study is to introduce a public key scheme based on LDLCs to overcome the drawbacks of the GGH cryptosystem. We use the properties of lattices and coding theory to improve the efficiency and security of our scheme. By exploiting an efficient iterative LDLC decoding algorithm, the decryption complexity is reduced. Moreover, by performing the Hermite normal form (HNF) [8] of the generator matrix of the used LDLC, the public key length is decreased. To improve the security level, we use the Gaussian vector whose variance is upper bounded with Poltyrev limit [9] as the perturbation vector. In this case, the proposed scheme can resist the most efficient attack to GGH cryptosystem, called as Embedding attack [6]. Also, by employing the Fujisaki-Okamoto conversion [10], the proposed scheme is not insecure against CCA2 and broadcast attack.

This paper is organized as follows. Section II gives an introduction to lattices, low density lattice codes and also reviews the construction of GGH cryptosystem. Section III explains the concept of the proposed public key cryptosystem based on LDLCs. The efficiency and security of the proposed cryptosystem are also assessed in Sections IV and V, respectively. Finally, Section VI concludes the paper.

## II. LATTICES, LOW DENSITY LATTICE CODES AND GGH CRYPTOSYSTEM

### A. Lattices and Low Density Lattice Codes

An $n$-dimensional *lattice* $\mathcal{L}$ over $\mathbb{R}^m$, where $n \leq m$, is the set of all integer linear combinations of some linearly independent vectors $\boldsymbol{g}_1, \cdots, \boldsymbol{g}_n \in \mathcal{L}$, i.e., $\mathcal{L} = \{\sum_i k_i \boldsymbol{g}_i | k_i \in \mathbb{Z}\}$. The vectors $\boldsymbol{g}_1, \cdots, \boldsymbol{g}_n$ are called *basis vectors* and the set of vectors $\{\boldsymbol{g}_1, \cdots, \boldsymbol{g}_n\}$ is called a *basis* of $\mathcal{L}$. In this case, the dimension of lattice $\mathcal{L}$ is $n$ and its *basis matrix* (also called *generator matrix* in channel coding) is the $n \times m$ matrix $G$ whose $n$ rows are the basis vectors $\boldsymbol{g}_1, \cdots, \boldsymbol{g}_n$. The lattice constructed by the generator matrix $G$ is the set of all integral linear combinations of the rows of $G$, i.e., $\mathcal{L}(G) = \{G\boldsymbol{v}: \boldsymbol{v} \in \mathbb{Z}^n\}$, where $\boldsymbol{v}$ is an $n$-dimensional vector of integers. Hence, every point of the lattice $\mathcal{L}(G)$ is of the form $\boldsymbol{x} = G\boldsymbol{v}$. The $i$-th successive minima of lattice $\mathcal{L}(G)$, denoted by $\lambda_i(\mathcal{L}(G))$, is the smallest real number such that there exist $i$ non-zero linear independent vector $\boldsymbol{v}_1, \cdots, \boldsymbol{v}_i \in \mathcal{L}(G)$ with $\|\boldsymbol{v}_1\|, \cdots, \|\boldsymbol{v}_i\| \leq \lambda_i(\mathcal{L}(G))$. The lattice $\mathcal{L}(G)$ is called *full-dimensional* if $n = m$ [7, 11]. In the rest of this paper, for simplicity, we assume that $\mathcal{L}(G)$ is full-dimensional and its generator matrix is the $n \times n$ square matrix. However, this case can be easily extended to the non-square generator matrix. Let $G$ be a non-singular $n \times n$ matrix, the *orthogonality defect* and *dual orthogonality defect* of $G$ are obtained as $\text{odf}(G) = |\det(G^{-1})| \prod_i \|\boldsymbol{g}_i\|$ and $\text{odf}^*(G) = |\det(G)| \prod_i \|\boldsymbol{g}_i^*\|$, respectively where, $\boldsymbol{g}_i^*$ is the $i$-th row in $G^{-1}$. One well-known equivalent lower triangular form related to the generator matrix $G$ is called the *Hermite normal form* (HNF) of $G$, denoted by $G' = \text{HNF}(G)$. It is shown that given an $n \times n$ nonsingular integer valued matrix $G$, there exist an $n \times n$ unimodular matrix $U$ such that the HNF of $G$ is obtained as $G' = \text{HNF}(G) = UG$ [8]. In this case, the entries of lower triangular $G' = [G'_{i,j}]$, $\forall 1 \leq i, j \leq n$, have the following properties: (1) if $i < j$, then $G'_{i,j} = 0$ (2) if $i = j$, then $G'_{i,j} \geq 1$ (3) if $i > j$, then $0 \leq G'_{i,j} < G'_{j,j}$.

The problem of finding a vector $\boldsymbol{v}_1 \in \mathcal{L}(G)$ such that $0 < \|\boldsymbol{v}_1\| \leq \lambda_1(\mathcal{L}(G))$ is called *shortest vector problem* (SVP) which is known as NP-complete problem [12]. Given a vector $\boldsymbol{w} \in \mathcal{L}(G)$, a basis $G$ and a vector $\boldsymbol{v}$ (which is usually not in the lattice), the problem of finding the lattice vector $\boldsymbol{w}$ closest to $\boldsymbol{v}$ is called *closest vector problem* (CVP). The complexity of general CVP was analyzed by van Emde Boas who showed that this problem is NP-hard [13]. In channel coding, CVP is referred to as *lattice decoding problem* (LDP) [11]. Another hard problem related to lattices is *shortest basis problem* (SBP) in which given a basis $B$ for a lattice in $\mathbb{R}^n$, the goal is to find the basis $B'$ which has the smallest orthogonality defect. It is known that there is no known polynomial time algorithm to solve SBP [2]. An $n$-dimensional lattice code over $\mathbb{R}^n$ is defined by an $n$-dimensional lattice $\mathcal{L}(G) \subset \mathbb{R}^n$ and a shaping region $\mathcal{B} \subset \mathbb{R}^n$ [11, 14]. In this case, the codewords are all the points of lattice $\mathcal{L}(G)$ which are lied within the shaping region $\mathcal{B}$. Inspired by Low Density Parity Check (LDPC) codes and in the goal of finding practical low complexity lattice codes, Low Density Lattice Codes (LDLCs) are introduced [7]. An $n$-dimensional LDLC over $\mathbb{R}^n$ is an $n$-dimensional lattice code with a nonsingular generator matrix $G$, i.e., $\det(G) = 1$, for which the parity check matrix $H = G^{-1}$ is restricted to be sparse. The sparsity of $H$ in LDLCs is utilized to develop a linear complexity iterative decoding algorithm by which good error performance is attained at large block length. The $i$-th row degree, denoted by $r_i$, $i = 1, \cdots, n$, is the number of nonzero elements in the row $i$ of the parity check matrix $H$. Moreover, the $i$-th column degree, denoted by $c_i$, $i = 1, \cdots, n$, is defined as the number of nonzero elements in column $i$ of the parity check matrix $H$. An LDLC is regular if all the row/column degrees of $H$ are equal to a common degree $d$. A regular LDLC with degree $d$ is called Latin square LDLC if every row/column of the parity check matrix $H$ has the same $d$ nonzero values. In a Latin square LDLC, the values of the $d$ non-zero coefficients in each row and each column are some permutation of the values $h_1, h_2, \cdots, h_d$. The sorted sequence of these $d$ nonzero values $h_1 \geq h_2 \geq \cdots \geq h_d \geq 0$ is called the *generating sequence* of Latin square LDLC [7].

### B. The Structure of GGH Public Key Cryptosystem

The security of GGH public key cryptosystem [2] relies on the computational difficulty of CVP. To generate the key, two different bases, called as public basis (or public key) $B \in \mathbb{Z}^{n \times n}$ and private basis (or private key) $R \in \mathbb{Z}^{n \times n}$, of the same lattice $\mathcal{L} \subset \mathbb{R}^n$ are used. The private basis $R$ is an $n \times n$ matrix with a low dual orthogonality defect. It can be generated as $R = R' + kI$, where, $I$ is an $n \times n$ identity matrix, $R' = [R'_{ij}]$ is an $n \times n$ matrix such that $|R'_{ij}| \leq l$ and $k \approx \sqrt{n}l$ for some constant $l$. The public key $B$ is an $n \times n$ matrix with a high dual orthogonality defect such that generates the same lattice as $R$, i.e., $\mathcal{L}(R) = \mathcal{L}(B)$. The public key $B$ is generated as $B = RU_1.U_2.\cdots$, where $U_i$ is the unimodular matrix (a matrix with integer entries and determinant of unit magnitude). To encrypt a message $\boldsymbol{m} \in \mathbb{Z}^n$, the ciphertext is obtained as $\boldsymbol{c} = B\boldsymbol{m} + \boldsymbol{e}$. The vector $\boldsymbol{e} = \{\pm\beta\}^n$ is called perturbation (or error) vector, where, $\beta$ is a small constant. At the receiver, the vector $\boldsymbol{v} = B\boldsymbol{m}$ is recovered as $\boldsymbol{v} = T\lceil R^{-1}\boldsymbol{c}\rfloor$, where $T = B^{-1}R$ is the unimodular matrix and $\lceil R^{-1}\boldsymbol{c}\rfloor$ denotes the vector obtained by rounding each entry in $R^{-1}\boldsymbol{c}$ to the nearest integer. Hence, $\lceil R^{-1}\boldsymbol{c}\rfloor = T\lceil R^{-1}(RT^{-1}\boldsymbol{v} + \boldsymbol{e})\rfloor = \boldsymbol{v} + T\lceil R^{-1}\boldsymbol{e}\rfloor$ and the decryption works if $\lceil R^{-1}\boldsymbol{e}\rfloor = 0$.

## III. THE PROPOSED PUBLIC KEY SCHEME

### A. Why We Use LDLCs?

At this point, the main question is that why we employ LDLCs in the structure of proposed public key scheme. To response this question, we cite the following reasonable causes: (1) One important subject about the proposed public key scheme is to find the families of easily and low complexity decodable lattice codes for a legitimate receiver (by the knowledge of private key), such that the decoding of these codes is infeasible for an active attacker (without the knowledge of private key) in polynomial time. It is shown that LDLCs have efficient and low complexity iterative decoding algorithm by which the better error performance can be attained [7] compared to other similar codes such as; LDPC lattice codes [15]. Moreover, if the HNF of their generator matrix is considered as the public key, then the decoding of these codes is infeasible for the active attacker. (2) LDLCs have a sufficient tradeoff between error performance and code length. This property allows us to use larger values of security parameters in the proposed scheme which leads to be secure against the cryptanalytic attacks, such as; the round-off attack and the nearest plane attack (see Sec. V). (3) By exploiting the properties of LDLCs, the security of the proposed scheme is based on two inherent hard lattice problems, i.e., SBP and CVP. In fact, recovering the generator matrix $G$ from the public key $G' = \text{HNF}(G)$ is equivalent to solve SBP. Also, finding the closest lattice point to ciphertext $c$ requires solving CVP. This strategy leads to improve the security level of the proposed scheme. (4) By using the LDLCs in the proposed scheme, a relationship can be established between code-based cryptography and lattice-based cryptography. In this case, we can exploit from the efficient properties of codes and lattices, simultaneously to improve the security and efficiency.

*B. Key Generation*

The secret key set is $\mathcal{K}_{sec} = \{H, U\}$, where $H$ is the sparse parity check matrix of the used LDLC and $U$ is the $n \times n$ unimodular matrix. The umimodular matrix $U$ is used for converting the generator matrix $G = H^{-1}$ of the used LDLC to its Hermite normal form (HNF). To obtain the public key, denoted by $G'$, first the generator matrix of the used LDLC is constructed as $G = H^{-1}$ and then its HNF, i.e., $G' = \text{HNF}(G) = UG$, is considered as the public key. In this case, the generator matrix of the used LDLC and its HNF are two different bases of the same lattice, i.e., $\mathcal{L}(G') = \mathcal{L}(G)$. The HNF $G'$ depends on the lattice $\mathcal{L}(G')$ not on the generator matrix $G$. Therefore, the public key $G'$ gives no information about the parameters of the secret key set, i.e., $H = G^{-1}$ and $U$.

*C. Encryption*

In the proposed public key scheme, an integer information sequence is considered as a plaintext vector (message) $m \in \mathbb{Z}^n$. To encrypt the message $m$, the sender (Alice) first fetches the public key $G' = \text{HNF}(G)$ from the public directory. Then, she constructs a codeword (lattice point) as $x = mG' = m'G$, where $m' = mU$. Finally, the ciphertext is obtained as $c = mG' + e = x + e$, where $e = (e_1, \cdots, e_n) \sim \mathcal{N}(\mu, \sigma^2)$ is a Gaussian vector with mean $\mu$ and variance $\sigma^2 < 1/(2\pi e)$. The ciphertext $c$ is transmitted through the insecure noiseless channel. Due to addition of codeword $x$ and Gaussian perturbation vector $e$, the transmission of ciphertext $c$ through noiseless channel is equivalent to transmission of codeword $x$ through AWGN channel without power restrictions. It is shown that for power unconstrained AWGN channel, there exists a lattice code of dimension $n$ such that the lattice point (codeword) $x = mG'$ can be decoded with arbitrarily small error probability if and only if the variance $\sigma^2$ of noise vector $e$ is less than $\sigma^2_{max} \triangleq$ ($\sqrt[n]{|\det(G)|^2})/(2\pi e)$ [7, 9]. In fact, this maximum noise variance that a lattice code should tolerate to have reliable communication over unconstrained AWGN channel is called *Poltyrev limit* [9]. In the proposed scheme, since the generator matrix of the used LDLC is non-singular, i.e., $\det(G) = 1$, the maximum noise variance is equal to $\sigma^2_{max} = 1/(2\pi e)$. Therefore, to achieve the reliable communication and also have small error probability, we consider the Gaussian vector with variance $\sigma^2 < 1/(2\pi e)$ as the perturbation vector $e$.

*D. Decryption*

The legitimate receiver (Bob) observes the ciphertext $c = x + e$. In this case, by the knowledge of secret key set $\mathcal{K}_{sec}$, he constructs the generator matrix $G = H^{-1}$ of the used LDLC and attempts to estimate the closest lattice point to the ciphertext $c$, i.e., $\hat{x} = \widehat{m'}G$, using the low complexity iterative LDLC decoding algorithm. The estimate of lattice point $\hat{x}$ is not directly found in the LDLC decoding algorithm. Instead, the probability density function (pdf) of the codeword $x = (x_1, \cdots, x_n)$, i.e., $\hat{f}_{x_i|c}(x|c)$, $i = 1, 2, \cdots, n$, is estimated. In fact, the $i$-th element of codeword $x$, denoted by $x_i$, is estimated as $\hat{x}_i = \text{argmax} \hat{f}_{x_i|c}(x|c)$ and hence the estimated codeword $\hat{x} = (\hat{x}_1, \cdots, \hat{x}_n)$ can be obtained [7]. Then, the estimation of the vector $m'$ is obtained as $\widehat{m'} = \widehat{m}U = [\hat{x}G^{-1}] = [\hat{x}H]$. Finally, the message is estimated as $\widehat{m} = \widehat{m'}U^{-1}$. Fig. 1 illustrates the block diagram of the proposed public key scheme based on LDLCs. As can be viewed from this figure, at the first step, Alice fetches $G'$ from the public directory. Then, she constructs the codeword (lattice point) as $x = m'G$ and adds the Gaussian perturbation vector with zero-mean and variance $\sigma^2 < 1/(2\pi e)$ to it. Also, the ciphertext $c = x + e$ is transmitted over the insecure noiseless channel. On the other hand, Bob, by the help of LDLC decoding algorithm, can estimate the closest lattice point to the ciphertext, i.e., $\hat{x} = \widehat{m'}G$, and recover $\widehat{m'} = [\hat{x}H]$. Finally, the message is estimated as $\widehat{m} = \widehat{m'}U^{-1}$. It is clear that the encryptor and the LDLC encoder are combined together at the transmitter. Also, the decryptor and the LDLC decoder are joined together at the receiver.

Note that for the power constrained AWGN channel, the LDLC encoding/decoding operations must be performed with shaping region to prevent the codeword's power from being too large [7]. As aforementioned in Sec. III-C, since the variance of perturbation vector $e$ is bounded by Poltyrev limit, i.e., $\sigma^2 < 1/(2\pi e)$, the shaping region boundaries is ignored in the encoding/decoding algorithms of the used LDLC. This operation has the following advantages for our scheme: (1) Since the number of points of the used LDLC, without considering the shaping region, is usually huge for the active attacker (Oscar), he cannot use an exhaustive method to implement the LDLC decoding operation. In fact, recovering the generator matrix $G$ from the public key $G' = \text{HNF}(G)$ is equivalent to solve SBP for Oscar. Also, finding the closest lattice vector, i.e., $x' = m'G$, to the ciphertext $c$ is equivalent to solve LDP for him. (2) Ignoring the shaping region boundaries in the LDLC encoding /decoding can decrease the encryption/decryption complexity of the proposed scheme.

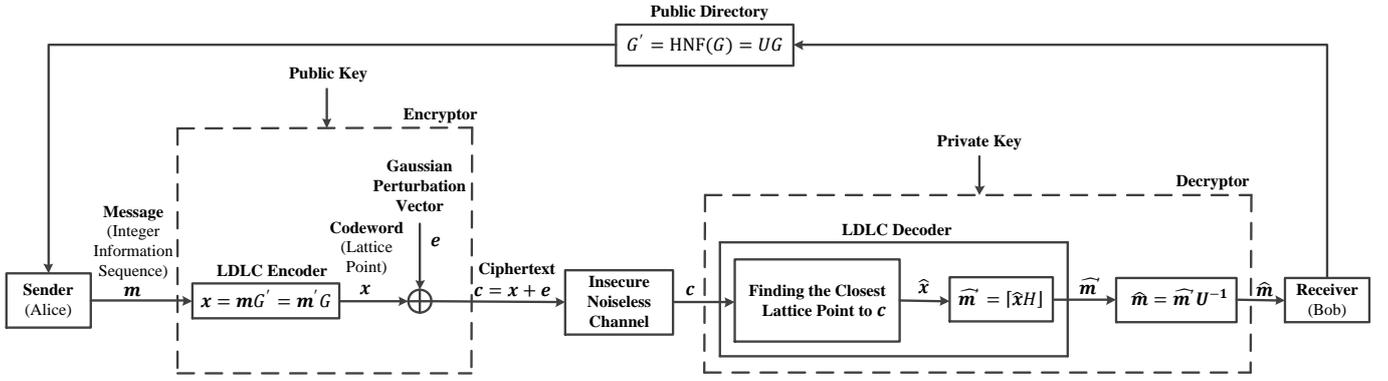

Fig. 1. The block diagram of the proposed public key cryptosystem based on LDLCs.

## IV. EFFICIENCY

In this section, the efficiency of the proposed cryptosystem is evaluated in terms of its key length and computational complexity.

### A. Key Length

It is shown that for large dimension $n \geq 1000$ and degree $d \leq 10$, a set of generating sequence $h_1, h_2, \cdots, h_d$, $h_1 = 1$, $\alpha \triangleq \sum_{i=2}^{d} h_i^2 / h_1^2 < 1$, can result the parity check matrix $H$ which satisfies all required properties to have efficient iterative LDLC decoding [7]. Moreover, the experiments on the security of the GGH-like cryptosystems based on HNF basis are shown that one necessary condition to achieve high security level is to apply the lattices with dimension $n \geq 782$ [16]. Hence, we use the LDLC with $n = 1000$ and degree $d \leq 10$ to improve the security and efficiency of the proposed scheme. We employ the Micciancio's idea [8] by which the public key of the proposed scheme is obtained as the HNF of the generator matrix of used LDLC. In this case, the public key size requires $\mathcal{O}(n^2 \log n)$ bits compared to GGH cryptosystem, whose public key size requires $\mathcal{O}(n^3 \log n)$ bits. Table I denotes the comparison between the public key lengths of the GGH and the proposed schemes. It is clear that applying the large dimensions, i.e., $n = 700, 1000, 1500$, for the original GGH scheme is impractical. However, the key length of the proposed scheme, by using the Micciancio's method is decreased significantly compared to the key lengths of GGH public key scheme. For example, for the dimension equal to 1000, the key length of the proposed scheme decreases up to 97 percent.

TABLE I
COMPARING THE KEY LENGTHS OF THE GGH AND PROPOSED SCHEMES.

| Scheme / Dimension | GGH Scheme | Proposed Scheme |
|---|---|---|
| $n$ | $\mathcal{O}(n^3 \lg n)$ | $\mathcal{O}(n^2 \lg n)$ |
| 700 | 14.18MB | 468KB |
| 1000 | 43.6MB | 1MB |
| 1300 | 99.4MB | 1.77MB |
| 1500 | 151.3MB | 2.4MB |

### B. Computational Complexity

As illustrated in Fig. 1, encryption procedure is performed by calculating the codeword $x = mG'$ and then adding the Gaussian perturbation vector $e$ to $x$. Therefore, the encryption complexity can be expressed as $C_{Enc} = C_{enc}(LDLC) + C_{add}(e)$, where, $C_{enc}(LDLC) = \mathcal{O}(n^2)$ is the encoding complexity of the used LDLC and $C_{add}(e) = \mathcal{O}(n \lg m)$ is the number of required binary operations for addition of $n$-symbol Gaussian perturnbation vector $e$ to the codeword $x$ (by asuming that the entries of $e$ are bounded by $m$). However, in the case of implementing a CCA2-secure variant (see Sec. V-D), the complexity of performing some suitable scrambling operation on the message $m$ before multiplication by $G'$ should be considered. The decryption complexity of this scheme is expressed as $C_{Dec} = C_{dec}(LDLC) + C_{mul}(\widehat{m'} U^{-1})$, where $C_{dec}(LDLC) = \mathcal{O}(n)$ is the decoding complexity of the used LDLC and $C_{mul}(\widehat{m'} U^{-1}) = \mathcal{O}(n^2 \lg m)$ is the number of required binary operations to perform the product of the estimated vector $\widehat{m'} = [\widehat{x} H]$ by the inverse of unimodular matrix $U$ (by asuming that the entries of $U$ are bounded by $m$).

## V. SECURITY

In this section, the cryptanalytic strength of the proposed scheme against some well-known attacks is being examined.

### A. The Embedding Attack

The embedding attack [6] is an efficient method to directly solve $\gamma$-CVP and seems to be practically the best way to break a CVP-based public key cryptosystem. This attack has been successfully used by Nguyen [6] to break the GGH cryptosystem. To perform this attack against GGH scheme, first the integer vector $s = \{\beta\}^n$ is added to the ciphertext $c = Bm + e$, $e = \{\pm \beta\}^n$ of the GGH scheme and then the modular equation $c + s = mG' + e + s \equiv m_{2\beta} G'$ is obtained, where $m_{2\beta} = m \pmod{2\beta}$. By subtracting the vector $m_{2\beta} G'$ from the ciphertext $c$, the vector $c - m_{2\beta} G' = (m - m_{2\beta}) G' + e$ is obtained. Now, this equation is divided by $2\beta$, the equation $(c - m_{2\beta} G')/2\beta = m' G' + e/2\beta$ is obtained, where $m' = (m - m_{2\beta})/2\beta$. In this case, since the rational point $(c - m_{2\beta} G')/2\beta$ is known, the simplified CVP-instance with a much smaller perturbation vector $e/2\beta \in \left\{\pm \frac{1}{2}\right\}^n$ is obtained. The error vector length is now $\sqrt{n/4}$, compared to $\beta \sqrt{n}$, previously. By this way, the problem of decrypting ciphertexts (CVP-instances for which the error vector has entries $\pm \beta$) is reduced to a simpler CVP-instance for which the error vector has entries $\pm\frac{1}{2}$. As it is mentioned, the embedding attack is applicable if the employed perturbation vector $e$ has the particular form, e.g., $e = \{\pm\beta\}^n$. In the proposed public key scheme, since the perturbation vector $e$ is Gaussian vector with mean $\mu$ and variance $\sigma^2 < 1/(2\pi e)$, the entries of $e$ are not known exactly for attacker and therefore the modular equation $c + s = m_{2\beta}G'$ cannot be accessed. Hence, the simplified CVP-instance with a much smaller perturbation vector cannot be attained to break the proposed scheme.

*B. The Round-Off Attack*

In the round-off attack [2], first it is tried to multiply the inverse of public key $G' = \text{HNF}(G)$ by the ciphertext $c$ as $(G')^{-1}c = m + (G')^{-1}e$. Then, the attacker does an exhaustive search to find the vector $d = (G')^{-1}e$. If the vector $d$ is found by an exhaustive search, the message $m$ will be recovered successfully. Therefore, the size of the search space needed for finding the correct vector $d$ should be large enough to prevent the Round-off attack. Below, we evaluate an approximate size of the search space in the proposed public key scheme. Let $d_i$ and $e_i$ denote the $i$-th entry in the vectors $d = (G')^{-1}e$ and the perturbation vector $e$, respectively. Let $g''_i$ be the $i$-th row of $(G')^{-1}$ and $g''_{ij}$ be the $(i,j)$-th element in the matrix $(G')^{-1}$. Using these notations, we have $d_i = g''_i \cdot e = \sum_j g''_{ij} e_j$, $\text{E}[d_i] = 0$ and $\text{Var}[d_i] = \sum_j {g''_{ij}}^2 \text{E}[e_j^2] = (\sigma\|g''_i\|)^2$, where $\|g''_i\|$ is the Euclidean norm of the $i$-th row in $(G')^{-1}$. To calculate the size of search space for the vector $d$, it is assumed that each entry $d_i$ in $d$ is Gaussian and all the entries are independent. In this case, the size of search space is exponential in the differential entropy of the Gaussian random vector $d$. The differential entropy of $d$ with variance $\sigma^2$ is obtained as $h(d) = \frac{1}{2}\log(\pi e\sigma^2)$. Since it is assumed that $d_i$'s are independent, the differential entropy of the vector $d$ equals the sum of the differential entropies of the entries, i.e., $h(d) = \frac{n}{2}\log(\pi e\sigma^2) + \sum_i \log\|g''_i\|$. Hence, the size of search space is obtained as $\mathcal{N}_d = 2^{h(d)} = (\pi e)^{n/2} \cdot \sigma^n \cdot \prod_i \|g''_i\|$. In the proposed scheme, since $\sigma^2 < 1/(2\pi e)$, we have $\mathcal{N}_d < (1/2)^{n/2} \cdot \prod_i \|g''_i\|$. It is shown that the Round-off attack is capable to recognize the vector $d$ up to dimension 100. In the proposed scheme, since the used dimension is more than 1000, this attack is also doomed to fail.

*C. The Nearest Plane Attack*

In the nearest plane attack [2], a better approximation technique is used to improve the round-off attack. In this attack, by the knowledge of public key $G'$, it is tried to apply the LLL lattice reduction algorithm [17] to obtain a reduced basis $B = \{b_1, \cdots, b_n\}$. Now, for a given ciphertext $c$ and a reduced basis $B$, all the affine spaces are considered as $H_k = \{kb_n + \sum_{i=1}^{n-1} \alpha_i b_i : \alpha_i \in \mathbb{R}\}$ for all $k \in \mathbb{Z}$. Then, the hyperplane $H_k$ is found which is the closest point, denoted by $p$, to the ciphertext $c$. Also, the point $c - kb_n$ is projected onto the $(n-1)$-dimensional space which is spanned by $B' = \{b_1, \cdots, b_{n-1}\}$. By this way, a new point $c' = c - kb_n$ and a new basis $B'$ are obtained. Such algorithm is proceeded recursively to find the closest point to $c'$, denoted by $p'$, in this $(n-1)$-dimensional lattice $\mathcal{L}'$. Finally, by finding the point $p'$, the point $p = p' + kb_n$ can be computed. In summary, the nearest-plane attack is partitioned into two parts: (1) the offline process in which the public key $G' = \text{HNF}(G)$ is transformed to the reduced basis $B$ by using the lattice reduction algorithms; (2) the online process in which the inverse of reduced basis $B$ is multiplied by the ciphertext $c$ as in the similar way used in the Round-off attack. The work factor of this attack can be computed from the Euclidean norm of the rows in the generator matrix $G$. It is shown that the Nearest plane attack has a lower work factor than the Round-off attack. But its work factor grows exponentially with the dimension of the lattice. Experiments show that this attack is infeasible for the dimensions 140-150 [2]. For the proposed scheme, since the used dimension is more than 1000, this attack is also failed.

*D. The Adaptive Chosen Ciphertext Attack*

It is shown that both McEliece and GGH cryptosystems are insecure against adaptive chosen ciphertext attacks (CCA2) [18, 21]. Therefore, generic [10, 18] and specific [19] CCA2-secure conversions can be used to make these schemes secure against CCA2. All applied methods in these conversions are based on scrambling the message inputs. In the proposed public key scheme, the ciphertext is $c = mG' + e$, where $m$ is a message and $e$ is a Gaussian perturbation vector. In this case, if an adversary encrypts one of two messages $m_1$ and $m_2$ and obtains a ciphertext $c$, then the attacker can distinguish a plaintext $m_i$ if $\|m_iG' - c\| < \|m_jG' - c\|$. In such attack, given a ciphertext $c = mG' + e$, an adversary inputs the $c + m'G'$ to the decryption oracle for some $m'$ and obtains the output of the decryption oracle, i.e., $\widehat{m}$. Then, the adversary can find out the original message $m$ by calculating $m = \widehat{m} - m'$. It means that the proposed scheme is insecure against CCA2. To obtain the security against CCA2, we can apply the Fujisaki-Okamoto generic conversion [10] in the proposed scheme as follows. Let $\mathcal{E}_K$, $\mathcal{D}_K$ be a symmetric encryption and a decryption form with a key $K$, respectively. Let $E$ and $F$ be random oracles. Then, $m' = E(e, m)$ and the ciphertext is obtained as $c = c_1 \| c_2 = (m'G' + e) \| \mathcal{E}_{F(e)}(m)$. For the decryption, using the LDLC decoding, first $\widehat{m''} = \widehat{m'}G'$ is estimated and $\widehat{m'} = \widehat{m''}(G')^{-1}$ is obtained. Then, the estimate of perturbation vector $\widehat{e} = c_1 - \widehat{m'}G'$ is computed. Now, a plaintext $\widehat{m} = \mathcal{D}_{F(e)}(c_2)$ with the secret key $F(e)$ is recovered. If $(E(\widehat{e}, \widehat{m}))G' + \widehat{e} = c_1$, then the output of decryption oracle is $\widehat{m}$. Otherwise, the decryption fails. By this way, an adaptive chosen ciphertext attack can be prevented for the proposed scheme.

*E. The Broadcast Attack*

In [20], Hastad presented an efficient attack, called as broadcast attack, in which the message is recovered in the broadcast scenario (i.e., sending a single message to different recipients using their respective public keys). Inspired by Hastad's method, two types of the broadcast attack are presented in [21] against the GGH cryptosystem. In these attacks, it is assumed that the sender encrypts the single message for different receivers. In this way, a random message $m$ is encrypted with different random public keys $G'_i, i =$ $1, \cdots, k$, is encrypted for different receivers and various ciphertexts $c_i = mG'_i + e, i = 1, \cdots, k$ are obtained. The goal is to recover the message $m \in \mathbb{Z}^n$ without any knowledge of receiver's secret key. It is shown that the intersecting lattice has interesting properties in the broadcast attack againt GGH scheme. Let $\mathcal{L}_1$ and $\mathcal{L}_2$ be two lattices, then the intersecting lattice of them is obtained as $\mathcal{L}_1 \cap \mathcal{L}_2 = (\mathcal{L}_1^* \cup \mathcal{L}_2^*)^*$. In the first type of broadcast attack, it is demonstrated that solving the shortest vector problem on the intersection lattice $\mathcal{L}_1 \cap \mathcal{L}_2$ will be easier than each of $\mathcal{L}_1$ or $\mathcal{L}_2$. In such attack, given the public keys $G'_i, i = 1, \cdots, k$, and the ciphertexts $c_i, i = 1, \cdots, k$, the embedding basis $G''_i = \begin{bmatrix} G'_i & 0 \\ c_i & 1 \end{bmatrix}$ is computed. Then, the intersection lattice $\mathcal{L}(G'') = \bigcap_{i=1}^{k} \mathcal{L}(G''_i)$ is obtained. Finally, the vector $(m\ 1)$ is found as the shortest vector of lattice $\mathcal{L}(G'')$. By this way, the message $m$ can be recovered. In the second type of broadcast attack, first, the basis $G'' = \sum_{i=1}^{k} G'_i$ is obtained. Then, the vector $c'' = \sum_{i=1}^{k} c_i$ is computed and the closest vector $v$ of $c''$ is found in the lattice $\mathcal{L}(G'')$. Finally, the message $m = v(G'')^{-1}$ is recovered.

To resist the proposed public key cryptosystem against broadcast attack, similar to CCA2-secure scheme, a random part (that is sufficiently big) should be added to the messages to prevent two messages to be equal under a reasonable probability. In fact, if the half of the message $m$ is random, the proposed scheme is secure against the broadcast attack. The scrambling method of message using the Fujisaki-Okamoto generic conversion for this scheme is suggested in Sec. V-D by which can be secure against this attack. However, using such conversion has repercussion in the computational complexity of this scheme.

## VI. CONCLUSION

The current paper was an attempt to address the issue of applying low density lattice codes in the structure of a public key cryptosystem. Compared to GGH cryptosystem, the efficiency of new scheme is improved from two following aspects: (1) due to the use of low complexity LDLC decoding, the decryption complexity is decreased; (2) because of considering the HNF of generator matrix as the public key, the key length is reduced up to 97 percent in the dimension $n = 1000$. Moreover, by exploiting the following strategies, the new scheme is better than the original GGH scheme from the security point of view: (1) using the Gaussian vector whose variance is upper bounded by Poltyrev limit, i.e., $\sigma^2 < 1/(2\pi e)$, as the perturbation vector to resist against the embedding attack; (2) increasing the dimension of used LDLC (due to the increased efficiency of the proposed scheme) to resist the round-off and the nearest plane attacks; (3) employing the generic Fujisaki-Okamoto conversion to resist against CCA2 and broadcast attacks.